\documentclass[sn-mathphys,Numbered]{sn-jnl}

\usepackage{graphicx}%
\usepackage{multirow}%
\usepackage{amsmath,amssymb,amsfonts}%
\usepackage{amsthm}%
\usepackage{mathrsfs}%
\usepackage[title]{appendix}%
\usepackage{xcolor}%
\usepackage{textcomp}%
\usepackage{manyfoot}%
\usepackage{booktabs}%
\usepackage{algorithm}%
\usepackage{algorithmicx}%
\usepackage{algpseudocode}%
\usepackage{listings}%
\usepackage{bm}

\usepackage{color}

\theoremstyle{thmstyleone}%
\theoremstyle{thmstyletwo}%

\theoremstyle{thmstylethree}%

\begin{document}

\title[Article Title]{Redundant basis interpretation of Doi-Peliti method and an application}


\author{\fnm{Shunta} \sur{Takahashi}}

\author*[]{\fnm{Jun} \sur{Ohkubo}$^*$}\email{johkubo@mail.saitama-u.ac.jp}

\affil{\orgdiv{Graduate School of Science and Engineering}, \orgname{Saitama University}, \orgaddress{\street{255 Shimo-Okubo, Sakura-ku}, \city{Saitama}, \postcode{338-8570}, \country{Japan}}}

\abstract{
The Doi-Peliti method is effective for investigating classical stochastic processes, and it has wide applications, including field theoretic approaches. Furthermore, it is applicable not only to master equations but also to stochastic differential equations; one can derive a kind of discrete process from stochastic differential equations. A remarkable fact is that the Doi-Peliti method is related to a different analytical approach, i.e., generating function. The connection with the generating function approach helps to understand the derivation of discrete processes from stochastic differential equations. Here, a redundant basis interpretation for the Doi-Peliti method is proposed, which enables us to derive different types of discrete processes. The conventional correspondence with the generating function approach is also extended. The proposed extensions give us a new tool to study stochastic differential equations. As an application of the proposed interpretation, we perform numerical experiments for a finite-state approximation of the derived discrete process from the noisy van der Pol system; the redundant basis yields reasonable results compared with the conventional discrete process with the same number of states.
}




\maketitle

\section{Introduction}

Stochastic processes are one of the most important research topics in physics, and many studies have proposed various techniques for stochastic processes to understand them and to evaluate statistics numerically. The Doi-Peliti method is well-known for investigating stochastic processes with chemical reaction types \cite{Doi1976a,Doi1976b,Peliti1985}. The Doi-Peliti method is based on the same algebraic procedure as the second-quantization method in quantum mechanics, and one can employ field-theoretic approaches such as perturbation methods and Feynman diagrams. Further extensions are possible; for example, some works discussed gene regulatory networks using spin operators in addition to the creation-annihilation operator in the Doi-Peliti method \cite{Sasai2003,Bhattacharyya2020}. We also apply the Doi-Peliti method to discuss duality relations in stochastic processes \cite{Ohkubo2010}; a stochastic differential equation could have a corresponding stochastic process with discrete states. The derivation is based on algebraic discussion, and the Doi-Peliti method is also applicable to derive it. Our purpose is not to give a non-exhaustive list of previous works; see, for example, review papers \cite{Tauber2005,Weber2017} and the text book \cite{Altland2010}.

The Doi-Peliti method is also related to other analytical approaches to stochastic processes, i.e., the Poisson representation and generating function approach \cite{Droz1994,Deloubriere2002, Weber2017}. For example, one can naively interpret the ket vector in the Doi-Peliti method with a monomial basis. Note that the creation-annihilation operators are a bit abstract, and there are several interpretations of the operators. It was clarified that certain orthogonal polynomials are available to interpret the ket vectors on which the creation-annihilation operators act \cite{Ohkubo2012}. A recent work discusses an extension with $\mathfrak{su}(1,1)$ algebra \cite{Greenman2022}.

The Doi-Peliti method indicates that it is possible to derive a discrete-state process if a time-evolution operator for a system is expressed in terms of creation-annihilation operators. Of course, one cannot solve the derived discrete-state process analytically in general, and it is crucial to develop numerical methods to solve it. The discrete-state process has infinite states in principle, and one needs a finite-state approximation in practice. When the approximated finite-state process recovers the statistics of the original system, one could say that it grasps essential parts of the original one. As we wrote above, there are several interpretations of the ket vectors in the Doi-Peliti method. However, is there room for deriving various approximated processes?

In the present paper, we propose a new redundant basis interpretation. Although it is conventional to use orthogonal basis functions, a non-orthogonal basis is introduced here. We employ the Doi-Peliti method to evaluate statistics in stochastic differential equations numerically. In previous related works, the orthogonality of basis functions is crucial to derive the corresponding discrete-state process with the creation-annihilation operators. We will show that the redundant basis function needs a different derivation, which leads to a discrete-state process with long-range hopping. In addition, we need different types of interpretations for the bra vectors. Of course, the redundancy might be meaningless for an infinite-state system. However, if we consider finite-state approximations, the discrete-state process with the redundant basis functions can have merits to grasp the essence of the original stochastic differential equations. We give a numerical demonstration for a noisy van der Pol system; the derived discrete-state process with the redundant basis functions yields a reasonable estimation compared with the conventional discrete-state process with the same number of states.

This paper is organized as follows. In Sect.~2, we briefly review the previous studies. Section 3 presents the main proposition of this paper, i.e., the use of redundant basis functions. We also discuss the reasons why long-distance hopping occurs. Section 4 describes an application of the redundant basis functions; the derived discrete-state process can grasp the essence of evaluating the statistics of the original stochastic differential equation. Section 5 gives some concluding remarks.

\section{Background}

In this section, we briefly review the basics of the Doi-Peliti method.
The Doi-Peliti method has a few essential features for this work.

The main aim of this section is to show the connection between a stochastic differential equation and a corresponding discrete process. The Doi-Peliti method naturally connects them. In order to review it, we proceed with the discussion in reverse order. First, we rewrite a master equation for a chemical reaction type process using the creation-annihilation operators. We also review the conventional interpretation of the generating function approach and its application to derive a discrete-state process from a stochastic differential equation.

In addition, we will note that the Doi-Peliti method is open to several interpretations. In the current study, one of the interpretations based mainly on the generating function approach is employed. Even within this interpretation, we will use its variants for shifts of origin in Sect.~\ref{sec_application}. We will also give some comments on the interpretations.

\subsection{Master equations and Doi-Peliti method}
\label{sec_Doi_Peliti}

Consider the following master equation for a probability distribution $P(n,t)$ for the state $n$ at time $t$:
\begin{align}
\frac{d}{dt} P(n,t)
= \sum_{r} \left( c_r \alpha_r(n-v_r) P(n-v_r,t) - c_r \alpha_r(n) P(n,t) \right),
\label{eq_master_equation_one_variable_general}
\end{align}
where $c_r$ is the rate of the $r$-th event, $\alpha_r(n)$ is a state-dependent function related to the $r$-th event rate, and $v_r$ is a scalar value related to the state-change in the $r$-th event. It is helpful to see a concrete example, and here we consider the following simple one:
\begin{align}
\begin{cases}
\textrm{Event 1: } \,\, A \to A + A \quad  \big( n \to n+1  \quad \textrm{at rate} \,\,\, \gamma n  \big),\\
\textrm{Event 2: } \,\, A + A \to A \quad  \big( n \to n-1  \quad \textrm{at rate} \,\,\, \sigma^2 n (n-1) / 2 \big),
\end{cases}
\label{eq_sFKPP_naive_chemical_reactions}
\end{align}
where $\gamma$ and $\sigma$ are system parameters. The corresponding master equation is written as follows:
\begin{align}
\frac{d}{dt} P(n,t) 
= &\gamma(n-1)P(n-1,t) - \gamma n P(n,t) \nonumber \\
&+ \frac{\sigma^2}{2} n(n+1) P(n+1,t) - \frac{\sigma^2}{2} (n-1)n P(n,t).
\label{eq_sFKPP_naive_master_equation}
\end{align}
Hence, we obtain the form in Eq.~\eqref{eq_master_equation_one_variable_general} with 
\begin{align}
c_1 = \gamma, \,\, c_2 = \sigma^2/2, \,\, 
\alpha_1(n) = n, \,\, \alpha_2(n) = n(n-1), \,\,
v_1 = +1, \,\, v_2 = -1.
\end{align}

In the Doi-Peliti method, the creation and annihilation operators, $a^\dagger$ and $a$, are introduced. These operators satisfy the following bosonic commutation relation:
\begin{align}
\left[a, a^\dagger \right] \equiv a a^\dagger - a^\dagger a = 1, \qquad \left[ a, a \right] = \left[ a^\dagger, a^\dagger \right] = 0.
\end{align}
Each operator acts on a ket vector in a Fock space, $|n\rangle$, as follows:
\begin{align}
a^\dagger | n \rangle = | n+1 \rangle, \qquad a | n \rangle = n | n-1 \rangle,
\label{eq_action_operators_in_Doi_Peliti}
\end{align}
and the vacuum state is characterized as
\begin{align}
a | 0 \rangle = 0.
\end{align}
The corresponding bra vector, $\langle m |$, is introduced formally with the following inner product:
\begin{align}
\langle m | n \rangle = n! \delta_{m,n},
\label{eq_Doi_Peliti_orthogonality}
\end{align}
where $\delta_{m,n}$ is the Kronecker delta. This relation means that the ket vectors $\{| n\rangle\}$ are orthogonal to each other.

Using the operator formalism, the following time-evolution operator is introduced:
\begin{align}
\mathcal{L} (a^\dagger,a) = \sum_r c_r \left( \left(a^\dagger\right)^{l_r} - \left(a^\dagger\right)^{k_r} \right) a^{k_r},
\label{eq_time_evolution_operator_in_Doi_Peliti}
\end{align}
where $l_r, k_r \in \mathbb{N}_0$. Let us define the state at time $t$ in the Doi-Peliti method as follows:
\begin{align}
| P(t) \rangle \equiv \sum_{n=0}^\infty P(n,t) | n \rangle,
\label{eq_state_in_Doi_Peliti}
\end{align}
where $P(n,t)$ is equivalent to the probability distribution in Eq.~\eqref{eq_master_equation_one_variable_general}. Then, the time-evolution equation for the state $|P(t)\rangle$ is given by
\begin{align}
\frac{d}{dt} | P(t) \rangle = \mathcal{L} (a^\dagger,a) | P(t) \rangle.
\label{eq_time_ecolution_equation_in_Doi_Peliti}
\end{align}

For the simple example in Eq.~\eqref{eq_sFKPP_naive_chemical_reactions}, the time-evolution operator in Eq.~\eqref{eq_time_evolution_operator_in_Doi_Peliti} is denoted as
\begin{align}
\mathcal{L}(a^\dagger, a) = \gamma \left(\left(a^\dagger\right)^2 - a^\dagger\right) a + \frac{\sigma^2}{2} \left(a^\dagger - \left(a^\dagger\right)^2 \right) a^2.
\end{align}
Hence, we have $l_1 = 2, k_1 = 1, l_2 = 1, k_2 = 2$. We can recover the original master equation \eqref{eq_sFKPP_naive_master_equation} by simply comparing the coefficient of $| n \rangle$ on the left-hand side and that on the right-hand side. The action of $\langle n |$ on Eq.~\eqref{eq_time_ecolution_equation_in_Doi_Peliti} from the left also recovers it; note that the orthogonality in Eq.~\eqref{eq_Doi_Peliti_orthogonality} is crucial.

We here give some comments. Firstly, it is straightforward to extend the above discussion to multivariate cases; the master equation for the multivariate cases is written as
\begin{align}
\frac{d}{dt} P(\bm{n},t)
= \sum_{r} \left( c_r \alpha_r(\bm{n}-\bm{v}_r) P(\bm{n}-\bm{v}_r,t) - c_r \alpha_r(\bm{n}) P(\bm{n},t) \right),
\end{align}
where $\bm{v}_r$ is not a scalar but a vector related to the state change for the $r$-th event. We note that the multivariate cases are also important to consider field-theoretic approaches; a discrete lattice on a space leads to creation-annihilation operators on each lattice point. Hence, the sub-indices for the creation-annihilation operators should be introduced as $a_i^\dagger$ and $a_i$. However, in the following discussions, we mainly use one-variable cases because of the notational simplicity. Secondly, note that it is possible to formally write the state $|P(t)\rangle$ as
\begin{align}
| P(t) \rangle = \exp\left[ \int_0^T \mathcal{L} (a^\dagger,a) dt\right] | P(0) \rangle,
\label{eq_Doi_Peliti_formal}
\end{align}
where $|P(0)\rangle$ is the initial state at time $t=0$. This formal expression of the solution $|P(t)\rangle$ is helpful in the following discussions.

\subsection{Doi-Peliti method and generating function approach}

There are some different approaches to classical master equations, as denoted in Sect.~1. The generating function approach is widely used, and it has been known that there is a connection between the Doi-Peliti method and the generating function approach; see the review \cite{Weber2017} for details.

In the generating function approach, firstly we define a generating function $G(x,t)$ as follows:
\begin{align}
G(x,t) = \sum_{n=0}^\infty P(n,t) x^n,
\end{align}
where $x \in \mathbb{R}$ is an additional variable. Note that the coefficients, $\{P(n,t)\}$, correspond to the probability distribution of the original system, and hence it is possible to derive the time-evolution equation for $G(x,t)$:
\begin{align}
\frac{d}{dt} G(x,t) = \mathcal{L}\left(x,\frac{d}{dx}\right) G(x,t),
\label{eq_time_evolution_generating_function}
\end{align}
where $\mathcal{L}\left(x,\frac{d}{dx}\right)$ is the time-evolution operator in terms of $x$ and $d/dx$. For example, the time-evolution operator for the simple example in Eq.~\eqref{eq_sFKPP_naive_chemical_reactions} is given as
\begin{align}
\mathcal{L}\left( x, \frac{d}{dx} \right) 
= \gamma (x^2-x) \frac{d}{dx} + \frac{\sigma^2}{2}(x-x^2) \frac{d^2}{dx^2}.
\end{align}
It is straightforward to recover the original master equation in Eq.~\eqref{eq_sFKPP_naive_master_equation} by comparing the coefficients of $x^n$ on both sides in Eq.~\eqref{eq_time_evolution_generating_function}.

Here, one can clearly see the following correspondence with the Doi-Peliti method and the generating function approach:
\begin{align}
a^\dagger \leftrightarrow x, \qquad a \leftrightarrow \frac{d}{dx}.
\label{eq_interpretation_with_generating_function_approach}
\end{align}
Actually, by using the interpretation of the ket vector $|n\rangle$ as
\begin{align}
| n \rangle \leftrightarrow x^n,
\label{eq_ket_generating_function}
\end{align}
we recover the actions of the creation and annihilation operators in Eq.~\eqref{eq_action_operators_in_Doi_Peliti} adequately.
A concrete interpretation for the bra vectors is given as follows \cite{Ohkubo2010}:
\begin{align}
\left\langle n \right\rvert \equiv
\int  dx \, \delta(x) \left( \frac{d}{dx} \right)^n (\cdot),
\end{align}
which yields the orthogonal relation in Eq.~\eqref{eq_Doi_Peliti_orthogonality}.

Note that there is no need to use concrete interpretations of the creation-annihilation operators when one employs the Doi-Peliti method with the field-theoretic approach such as Feynman diagrams. However, as we will see below, the interpretation with the generating function approach is beneficial to connect stochastic differential equations and discrete-state processes.

There are some comments on the interpretations of the Doi-Peliti method. The above discussion is based on the generating function approach; for example, it connects the annihilation operator with the differential operator. However, it is possible to consider different interpretations; for example, we can employ the Hermite polynomials and Charlier polynomials as the interpretation \cite{Ohkubo2012}. In Sect.~\ref{sec_application}, we will use a slightly different interpretation from Eqs.~\eqref{eq_interpretation_with_generating_function_approach} and \eqref{eq_ket_generating_function}.

\subsection{Stochastic differential equations and derived discrete processes}
\label{sec_review_dual}

A stochastic differential equation with one variable is denoted as follows:
\begin{align}
dX = a(X) dt + b(X) dW(t),
\label{eq_sde_one_variable}
\end{align} 
where $X \in \mathbb{R}$ is a stochastic variable, $a(x)$ is a drift coefficient function, and $b(x)$ is a diffusion coefficient function. $W(t)$ is a Wiener process, which corresponds to the noise of the system. See, for example, \cite{Gardiner2009} for details of the stochastic differential equations.

It is known that some stochastic differential equations have corresponding discrete-state stochastic processes. That is, the duality in stochastic processes connects them; for example, see \cite{Shiga1986,Giardina2007,Giardina2009,Carinci2013,Carinci2015}. The discrete-state process is sometimes easily solved analytically, and the solution is beneficial to obtain the statistics in stochastic differential equations. Recent studies clarified that the Doi-Peliti method and related approaches are beneficial to derive the dual process \cite{Ohkubo2010,Ohkubo2013,Greenman2020}. Here, a brief explanation is given with a simple example. Although it is possible to discuss the connection between the stochastic differential equation and the discrete-state process only in terms of the Doi-Peliti method, the following review is based on the generating function approach \cite{Ohkubo2019}.

Consider the following example:
\begin{align}
dx = - \gamma x(1-x) dt + \sigma \sqrt{x(1-x)} dW(t)
\label{eq_sFKPP}
\end{align}
for $0 \le x \le 1$, where $\gamma$ and $\sigma$ are parameters. This system is related to the stochastic Fisher-Kolmogorov-Petrovsky-Piscounov (sFKPP) equation without a space-diffusion term \cite{Panja2004}.

It is well-known that the stochastic differential equation in Eq.~\eqref{eq_sde_one_variable} has the corresponding Fokker-Planck equation which describes the time-evolution of the density function $p(x,t)$ \cite{Gardiner2009,Risken1989}:
\begin{align}
\frac{\partial}{\partial t} p(x,t) &=
- \frac{d}{dx} a(x) p(x,t) + \frac{1}{2} \frac{d^2}{dx^2} \left( b(x) \right)^2 p(x,t) \nonumber \\
&= \mathcal{L}^\dagger p(x,t),
\end{align}
where $\mathcal{L}^\dagger$ is the time-evolution operator. For the sFKPP equation, we have
\begin{align}
\mathcal{L}^\dagger
= - \frac{\partial}{\partial x} 
\left[ - \gamma x(1-x) \right]
+ \frac{1}{2} \frac{\partial^2}{\partial x^2}
\left[ \sigma^2 x(1-x)  \right].
\label{eq_sFKPP_Fokker_Planck}
\end{align}

The statistics in the stochastic differential equation are evaluated with the density function $p(x,t)$. The famous statistic is the $m$-th moment, which is defined as
\begin{align}
\mathbb{E}_T \left[x^m \right] = \int_{-\infty}^\infty dx \,  x^m  p(x,T),
\end{align}
where $\mathbb{E}_t[\cdot]$ means the expectation at time $t$. Using the integration by parts, the calculation of the moment is rewritten as follows:
\begin{align}
\mathbb{E}_T[x^m]
&= \int_{-\infty}^\infty dx \, x^m \left( e^{\int_0^T \mathcal{L}^\dagger dt } p(x,0) \right) \nonumber \\
&= \int_{-\infty}^\infty dx \, \left( e^{\int_0^T \mathcal{L} dt }  x^m \right) p(x,0) \nonumber \\
&= \int_{-\infty}^\infty dx \, \varphi^{(m)}(x,T) p(x,0), 
\label{eq_duality}
\end{align}
where $\mathcal{L}$ is the adjoint operator of $\mathcal{L}^\dagger$. Note that $p(x,T)$ is the density function so that $p(x,T) \to 0$ when $x \to \pm \infty$. In the final expression, $\varphi^{(m)}(x,T)$ is given as the time-evolution of $x^m$ with the time-evolution operator $\mathcal{L}$. 

Here, let us focus on Eq.~\eqref{eq_Doi_Peliti_formal} and Eq.~\eqref{eq_ket_generating_function}; formally, $\varphi^{(m)}(x,T)$ is the solution of the time-evolution equation for the state in the Doi-Peliti method. Actually, it is possible to derive a discrete-state process corresponding to the time-evolution of $\varphi^{(m)}(x,T)$. For the sFKPP equation in Eq.~\eqref{eq_sFKPP}, it is clarified that the discrete-state process is given as Eq.~\eqref{eq_sFKPP_naive_chemical_reactions}. By evaluating the probability distribution $\{P(n,T)\}$ at time $T$, $\varphi^{(m)}(x,T)$ is adequately calculated, i.e.,
\begin{align}
\varphi^{(m)}(x,T) = \sum_{n=0} P(n,T) x^n,
\end{align}
which immediately yields the statistics of the original stochastic differential equation via Eq.~\eqref{eq_duality}. Of course, the initial condition for $\{P(n,0)\}$ should be $P(n,0) = \delta_{n,m}$, which corresponds to the evaluated moment in Eq.~\eqref{eq_duality}.

As seen above, the interpretation with the generating function approach and the Doi-Peliti method yields the discrete-state process naturally. Here, we briefly mention the multivariate case. For the multivariate stochastic differential equation,
\begin{align}
\mathrm{d}\bm{X} = \bm{a}(\bm{X})\mathrm{d}t + B(\bm{X})\mathrm{d}\bm{W}(t),
\end{align} 
we have the following multivariate Fokker-Planck equation:
\begin{align}
\frac{\partial}{\partial t}p(\bm{x},t) = 
 -\sum_{i}\frac{\partial}{\partial x_i}a_i(\bm{x}) p(\bm{x},t) + \frac{1}{2}\sum_{i,j}\frac{\partial^2}{\partial x_i\partial x_j}[B(\bm{x})B(\bm{x})^{\mathrm{T}}]_{ij} p(\bm{x},t),
\end{align}
where $\bm{a}(\bm{x})$ is a vector-valued function, and $a_i(\bm{x})$ is the $i$-th element; $B(\bm{x})$ is a matrix-valued function related to the noise term. In the multivariate case, the Doi-Peliti method should also be extended to the multivariate one, as discussed above. 

The important comment here is the stochasticity of the derived discrete-state process. For the sFKPP equation, the derived discrete-state process is a stochastic process. Note that there is no guarantee in general that the derived process has stochasticity; actually, it is rare to have the derived discrete-state process with stochasticity. However, it is possible to recover the stochasticity by extending the process \cite{Ohkubo2013}. In addition, there is no need to pay attention to stochasticity in the following discussions. Hence, we omit the recovery of the stochasticity in the remaining part. Then, we will use the notation, `the derived discrete-state process,' in the following sections.

At the end of this section, we comment on the applicability of the above discussion. Because of the correspondence between the stochastic variable and the Doi-Peliti operators in Eq.~\eqref{eq_interpretation_with_generating_function_approach}, the above derivation of the discrete-state process is applicable when the coefficient functions are polynomials. Furthermore, as denoted in \cite{Ohkubo2013}, even if the coefficient functions include $\sin$ or $\cos$ functions, we can derive the corresponding discrete process by extending the stochastic variables. Although rigorous mathematical discussions remain still open, we can deal with various types of stochastic differential equations within the framework here. In addition, since the Doi-Peliti method has room for interpretation, the derived discrete-state process is not unique; in \cite{Ohkubo2019}, the Hermite polynomials give a different discrete-state process from Eq.~\eqref{eq_sFKPP_naive_master_equation}.

\section{Extension of derivation of discrete processes}

This section is one of the main contributions of this work; a redundant basis function is proposed. We will discuss the merits of the redundant basis function in the next section.

\subsection{Redundant basis functions}

Redundancy plays sometimes important roles. One of the famous examples is the wavelet (for example, see \cite{Starck2015}); the wavelet basis is redundant compared with the conventional Fourier basis functions, but it is beneficial to analyze signals and images. In the stochastic system, the Doi-Peliti method usually employs the orthogonal ket vector in Eq.~\eqref{eq_action_operators_in_Doi_Peliti}. As for a redundant basis in the Doi-Peliti method, coherent states for the creation-annihilation operators are famous, which lead to the path-integral formulation \cite{Weber2017}. However, the coherent states do not derive discrete-state processes. Is there any redundant basis that leads to discrete-state processes?

Considering the connection with the generating function, we here introduce redundant basis functions, $\{|\zeta_{n,d} \rangle\}$, where $n \in \mathbb{N}_0$. The index $d$ is an element in $\mathcal{S}$, where a set $\mathcal{S}$ is arbitrarily chosen depending on the redundancy. The basis satisfies the following properties for the creation-annihilation operators:
\begin{align}
a^\dagger |\zeta_{n,d} \rangle = \sum_{k} \eta^{\mathrm{(c)}}_{n,k} |\zeta_{k,d} \rangle,
\end{align}
and
\begin{align}
a |\zeta_{n,d} \rangle = \sum_{k} \eta^{\mathrm{(a)}}_{n,k} |\zeta_{k,d} \rangle,
\end{align}
where $\{\eta^{\mathrm{(c)}}_{n,k}\}$ and $\{\eta^{\mathrm{(a)}}_{n,k}\}$ are coefficients related to the creation and the annihilation, respectively. Of course, one might consider that the coefficients should also depend on $d$. However, the above setting is enough for our following discussions.

There are many candidates for the redundant basis  $|\zeta_{n,d} \rangle$, but we here only focus on the following one:
\begin{align}
|\zeta_{n,d} \rangle = x^n \exp\left( - \beta ( x - c_d )^2 \right),
\label{eq_gaussian_redundant_basis}
\end{align}
where $\beta$ is a common parameter, and $c_d \in \mathbb{R}$. 

Then, employing the interpretation of the generating function approach, i.e., Eq.~\eqref{eq_interpretation_with_generating_function_approach}, the following actions of the creation and annihilation operators are derived:
\begin{align}
a^\dagger |\zeta_{n,d} \rangle = x^{n+1} \exp\left( - \beta ( x - c_d )^2 \right) =  |\zeta_{n+1,d} \rangle
\label{eq_redundant_basis_with_creation}
\end{align}
and
\begin{align}
a |\zeta_{n,d} \rangle &= n x^{n-1} \exp\left( - \beta ( x - c_d )^2 \right) - 2\beta x^{n+1} \exp\left( - \beta ( x - c_d )^2 \right)  \nonumber \\
&\quad + 2 \beta c_d x^{n}  \exp\left( - \beta ( x - c_d )^2 \right) \nonumber \\
&= \, n |\zeta_{n-1,d} \rangle - 2\beta |\zeta_{n+1,d} \rangle  + 2 \beta c_d |\zeta_{n,d} \rangle.
\label{eq_redundant_basis_with_annihilation}
\end{align}
The redundant basis with the index $d$, $|\zeta_{n,d} \rangle$, is independent of the other index ones, $|\zeta_{n,d'} \rangle$ for $d' \neq d$. However, the different index cases are related to each other finally in the derived discrete-state process; we will explain it later.

Note that the basis is easily extended to multivariate cases:
\begin{align}
|\zeta_{n,d} \rangle = \bm{x}^{\bm{n}} \exp\left( - \beta \| \bm{x} - \bm{c}_d \|^2 \right),
\end{align}
where 
\begin{align}
\bm{x}^{\bm{n}} \equiv \prod_{i} x_i^{n_i}
\end{align}
and $\bm{c}_d$ is a vector.

\subsection{Derived discrete-state processes with long-range hopping}
\label{subsec_derivation_dual_process}

Different from the conventional ket vector, the redundant basis function does not satisfy the orthogonality, $\langle \zeta_{d'} | \zeta_d \rangle \neq 0$. Here, we write the inner product as
\begin{align}
\langle \zeta_{k,d'} | \zeta_{n,d} \rangle = \zeta_{(k,d'),(n,d)},
\label{eq_inner_product_for_Z}
\end{align}
where $\zeta_{(k,d'),(n,d)} \in \mathbb{R}$. If we rearrange the sub-indices of $\zeta_{(k,d'),(n,d)}$ so that $(k,d')$ is the row index and $(n,d)$ is the column one, it is possible to introduce a matrix $Z$ with
\begin{align}
Z = \left[\zeta_{(k,d'),(n,d)}\right].
\label{eq_definition_Z}
\end{align}

Note that there is no need to specify concrete forms of the bra and ket vectors in the case with the conventional ket vector; the formal settings in Sect.~\ref{sec_Doi_Peliti} are enough to derive a discrete process. However, in the case of the redundant basis, we should give specific concrete forms. Then, the generating function approach is helpful for the interpretation of the creation-annihilation operators. Additionally, for the redundant basis with Eq.~\eqref{eq_gaussian_redundant_basis}, we here introduce the corresponding bra vector as follows:
\begin{align}
\langle \zeta_{n,d} |
= \int dx \,  x^n \exp\left( - \beta ( x - c_d )^2 \right) \left( \cdot \right).
\end{align}
Then, the matrix elements of $Z$, i.e., the inner products of the bra and ket vectors, are explicitly evaluated; the moment-generating function of Gaussian distributions is beneficial to evaluate them numerically.

The non-orthogonality yields a complicated discrete-state process, as follows. Let us define a state $| \varphi^{(m)}(t) \rangle$ with the redundant basis:
\begin{align}
| \varphi^{(m)}(t) \rangle = \sum_{n,d} C(n,d,t) | \zeta_{n,d} \rangle,
\label{eq_proposed_ket_state_time_evolution}
\end{align}
where $\{C(n,d,t)\}$ are the expansion coefficients at time $t$. Then, we obtain the following time-evolution equation for the state $| \varphi^{(m)}(t) \rangle$ as in the conventional case:
\begin{align}
\frac{d}{dt} | \varphi^{(m)}(t) \rangle  = \mathcal{L}\left( a^\dagger, a\right) | \varphi^{(m)}(t) \rangle.
\end{align}

Different from the conventional Doi-Peliti method, the naive comparison of coefficients on both sides of the equation does not lead to the time-evolution equation for the coefficients $\{C(n,d,t)\}$ because of the lack of orthogonality in the basis. Hence, we must multiply the bra vector $\langle \zeta_{k,d'} |$ from the left of Eq.~\eqref{eq_proposed_ket_state_time_evolution}:
\begin{align}
\langle \zeta_{k,d'} | \left( \frac{d}{dt} \sum_{n,d}  C(n,d,t) | \zeta_{n,d} \rangle \right) 
= \langle \zeta_{k,d'} | \left( \sum_{n,d} \mathcal{L}\left( a^\dagger, a\right)  C(n,d,t)  | \zeta_{n,d} \rangle\right).
\end{align}
Then, we define the matrix elements of $\mathcal{L}\left( a^\dagger, a\right)$ with respect to the redundant basis as
\begin{align}
L_{(k,d'),(n,d)} = \left\langle \zeta_{k,d'} \left| \mathcal{L}\left( a^\dagger, a\right)  \right| \zeta_{n,d} \right\rangle.
\label{eq_definition_L}
\end{align}
Of course, this expression is formal, and the matrix size is infinite in general; we will employ a finite-state approximation later. Then, we obtain 
\begin{align}
\frac{d}{dt} \sum_{n,d} \zeta_{(k,d'),(n,d)} C(n,d,t) = \sum_{n,d} L_{(k,d'),(n,d)} C(n,d,t).
\end{align}
Finally, using the vector notation for the state coefficients,
\begin{align}
\bm{C}(t) = \left[C(n,d,t) \right],
\end{align}
we obtain the time-evolution equation for the derived discrete-state process as follows:
\begin{align}
\frac{d}{dt} \bm{C}(t) = Z^{-1} L  \bm{C}(t),
\label{eq_redundant_time_evolution_final}
\end{align}
where $Z^{-1}$ is the inverse of $Z$.

\begin{figure}[tb]
  \centering
  \includegraphics[width=12cm]{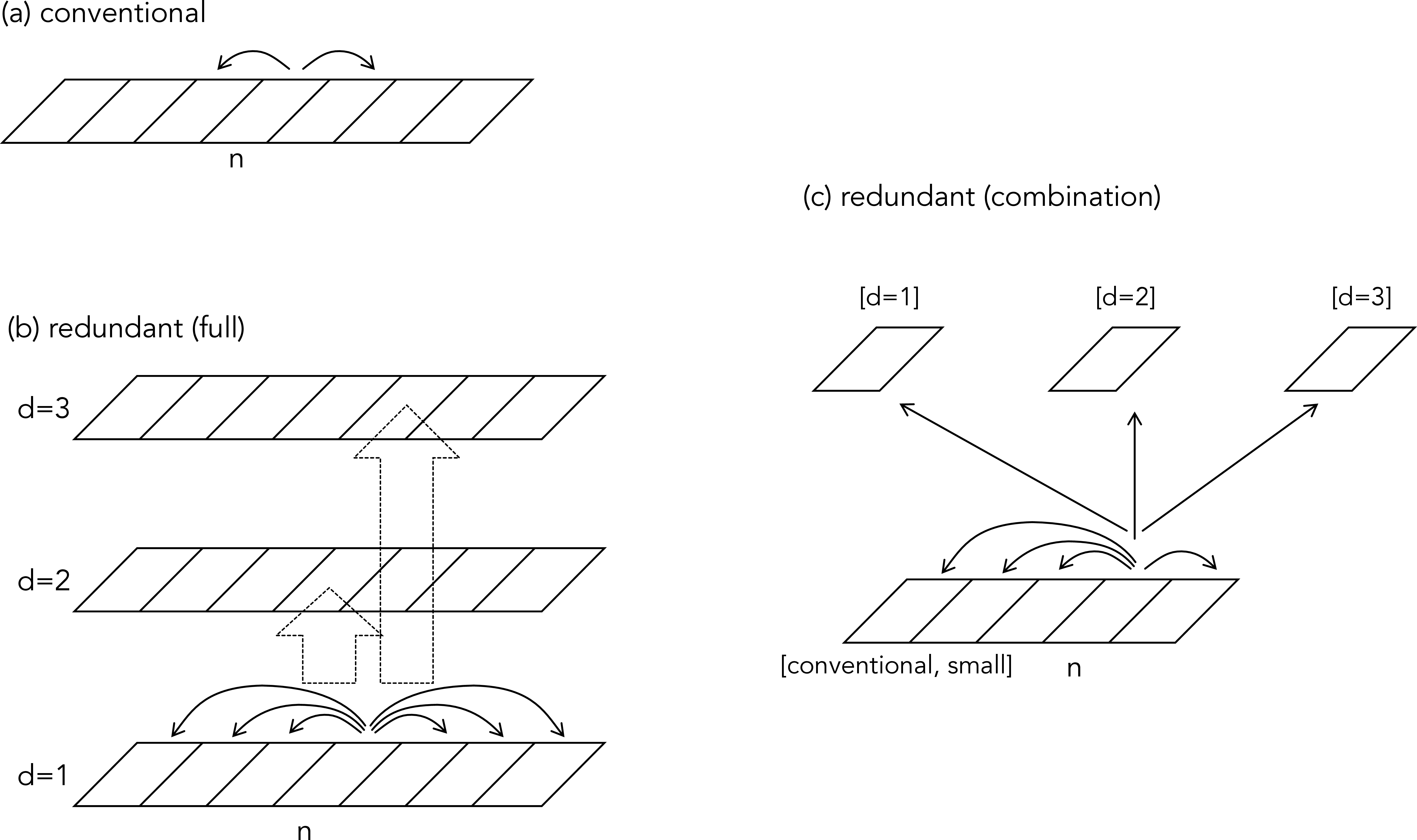}
  \caption{
Conceptual figures for derived processes. (a) The conventional basis yields a discrete-state process with local hopping. (b) The usage of the redundant basis functions causes not only long-range hopping but also hopping among different redundant states. (c) As demonstrated in Sec.~4, the finite approximation and combinations with the conventional and redundant basis give a process that captures an essential part of the original stochastic differential equation.
}
  \label{fig_discrete_process_from_redundant_basis}
\end{figure}

Recall that there is no interaction between basis functions with different $d$ in Eqs.~\eqref{eq_redundant_basis_with_creation} and \eqref{eq_redundant_basis_with_annihilation}. However, the existence of $Z^{-1}$ in Eq.~\eqref{eq_redundant_time_evolution_final} creates the interaction among states with different $d$. Furthermore, the existence of $Z^{-1}$ causes long-range hopping although there is only local hopping between neighboring states in the conventional Doi-Peliti method. Hence, we conclude that the redundant basis naturally causes various complicated processes. Figures~\ref{fig_discrete_process_from_redundant_basis}(a) and \ref{fig_discrete_process_from_redundant_basis}(b) give the conceptual pictures for these cases.

We summarize the points as follows:
\begin{itemize}
\item The conventional basis could yield a discrete-state process with only local hopping.
\item The redundant basis yields a discrete process not only with long-range hopping but also with hopping among different redundant states.
\end{itemize}

Of course, a different choice of redundant basis functions yields a different discrete-state process. Hence, it is clear that the derived discrete-state process is not unique; many discrete-state processes exist with a redundant basis. In addition, from the viewpoint of physics, one would have a question: Is there any merit in the redundant basis functions? The original stochastic differential equation and the derived discrete-state process have a kind of `dual' relation, and it will be interesting to seek what discrete-state process captures the essence of the original stochastic differential equation. As far as we know, the present work is the first one to show the existence of many discrete-state processes and their complex characteristics; a systematic survey is beyond the scope of the present paper, and further studies will be performed in the future. However, in the next section, we propose additional ideas for employing the redundant basis functions in applications.

\section{An application of derived process}
\label{sec_application}

In this section, we propose an application of the redundant basis functions. For the application, we combine the redundant basis functions with the conventional one, and the interpretation of the bra and ket vectors is modified a little with the aid of the generating function approach. Although practical studies are not the main purpose of the present work, we give a simple numerical demonstration; the redundant basis functions can capture the essence of the original stochastic differential equation correctly, compared with the case using only the conventional basis.

\subsection{Combinations of different-type discrete states}

In practical cases, it is impossible to consider discrete-state processes with infinite size. Then, one should use a kind of finite approximation. Note that the derived discrete-state process is employed to calculate the statistics in the original stochastic differential equation. Hence, the initial condition should have the same functional form as the statistics, as shown in Eq.~\eqref{eq_duality}. For simplicity, we consider a one-dimensional case. Then, we must evaluate the time-evolution equation $\varphi^{(m)}(x,t)$ in Eq.~\eqref{eq_duality}:
\begin{align}
\frac{d}{dt} \varphi^{(m)}(x,t) = \mathcal{L} \varphi^{(m)}(x,t),
\label{eq_time_evolution_phi}
\end{align}
with the initial condition
\begin{align}
\varphi^{(m)}(x,0) = x^m
\end{align}
when the $m$-th moment is evaluated.

As discussed above, we can interpret Eq.~\eqref{eq_time_evolution_phi} with the generating function approach, and hence it is connected to the Doi-Peliti method naturally. The proposal here is to use the following state $|\varphi^{(m)}(t)\rangle$ in the Doi-Peliti form:
\begin{align}
|\varphi^{(m)}(t)\rangle = \sum_{n \in \mathcal{S}_{\textrm{c}}} C^{\textrm{(c)}}(n,t) | n \rangle
+ \sum_{d\in \mathcal{S}_{\textrm{r}}} C^{\textrm{(r)}}(d,t) | \zeta_{0,d} \rangle,
\label{eq_propose_combined_state}
\end{align}
where $\mathcal{S}_{\textrm{c}}$ is the finite set of indices for the conventional basis functions, $\mathcal{S}_{\textrm{r}}$ is for the redundant basis ones. $\{C^{\textrm{(c)}}(n,t)\}$ and $\{C^{\textrm{(r)}}(d,t)\}$ are the corresponding coefficients. 
Note that the discussion in Sect.~\ref{subsec_derivation_dual_process} is a formal one because we assumed an infinite basis functions. Here, we use the finite number of basis functions. Hence, both the matrix $L$ in Eq.~\eqref{eq_definition_L} and the matrix $Z$ in Eq.~\eqref{eq_definition_Z} have finite sizes. Then, we can explicitly write the time evolution equation~\eqref{eq_redundant_time_evolution_final} for the discrete-state process.

This proposal means that we use the combination of the conventional basis and the redundant ones. Of course, it is possible to use other choices. However, the above one has the following two merits. The first one is that the redundancy is restricted because we use only $| \zeta_{n,d} \rangle$ with $n=0$. The second one is that the conventional basis is suitable to express the initial condition. Using the correspondence with the generating function approach, the initial condition, i.e., $x^m$, corresponds to $|m\rangle$. Hence, it is enough to set all coefficients zero except for $C^{\textrm{(c)}}(m,0) = 1$. Although it could be possible to approximate the initial condition with the linear combination of the redundant basis functions, we will demonstrate that the basis in Eq.~\eqref{eq_propose_combined_state} captures the characteristics of the discrete process adequately.

Note that the usage of the finite-dimensional basis functions yields a unique decomposition in general because the finite monomials cannot span the space derived from the Taylor expansion of exponential functions. Although the non-uniqueness property in the case of infinite-dimensional basis functions could be interesting from the viewpoint of mathematical physics, this is beyond the scope of the present paper. Here, we focus on only the finite-dimensional basis case.

For multivariate cases, we use 
\begin{align}
|\varphi^{(\bm{m})}(t)\rangle 
= \sum_{\bm{n} \in \mathcal{S}_{\textrm{c}}} C^{\textrm{(c)}}(\bm{n},t) | \bm{n} \rangle
+ \sum_{d\in \mathcal{S}_{\textrm{r}}}  C^{\textrm{(r)}}(d,t) | \zeta_{0,d} \rangle.
\end{align}

These discussion yields the derived discrete-state process. Figure~\ref{fig_discrete_process_from_redundant_basis}(c) depicts the proposal. In this case, we have long hopping on the conventional finite lattice, and the hopping to the redundant basis also exists. Of course, hopping among the redundant bases is also possible. Although intuitive explanations of the derived discrete-state process are beyond the scope of the present paper, it contains some essential parts to calculate the statistics in the original stochastic process, compared with the conventional discrete-state process. We finally demonstrate it next.

\subsection{Useful interpretation of operators}
\label{subsec_application}

Again, note that the aim here is to evaluate the statistics in the stochastic differential equation. In usual cases, the initial point in the stochastic differential equation is fixed at $x_\textrm{ini}$; this indicates that the initial probability density function has the following Dirac delta function:
\begin{align}
p(x,0) = \delta(x - x_\textrm{ini}).
\end{align}
Then, the $m$-th moment in Eq.~\eqref{eq_duality} becomes
\begin{align}
\mathbb{E}_T[x^m] = \int_{-\infty}^\infty dx \, \varphi^{(m)}(x,T) \delta(x - x_\textrm{ini}) = \varphi^{(m)}(x_\mathrm{ini},T),
\end{align}
which indicates that only the function value at $x_\mathrm{ini}$ is enough to evaluate the $m$-th moment. Hence, the following interpretations of the Doi-Peliti formalism with the generating function approach are employed:
\begin{align}
&a^\dagger \leftrightarrow (x-x_\mathrm{ini}), \label{eq_new_c}\\
&a \leftrightarrow \frac{d}{d(x-x_{\mathrm{ini}})}, \label{eq_new_a}\\
&| n \rangle = \left(x - x_{\mathrm{ini}}\right)^{n}, \label{eq_new_ket}
\end{align}
which correspond to the shift of origin. We also use the following redundant basis:
\begin{align}
| \zeta_{n,d} \rangle
= (x-x_\mathrm{ini})^n \left( \exp\left[ -\beta ( x - c_d )^2 \right] - \exp\left[ -\beta ( x_{\mathrm{ini}} - c_d )^2 \right] \right).
\label{eq_redundant_shift}
\end{align}
Note that the action of the annihilation operator $a$ to the redundant basis in Eq.~\eqref{eq_redundant_shift} is different from Eq.~\eqref{eq_redundant_basis_with_annihilation}. We finally obtain the following expression:
\begin{align}
\varphi^{(m)}(x,t)
= & \sum_{n \in \mathcal{S}_{\textrm{c}}} C^{\textrm{(c)}}(n,t) \left(x - x_{\mathrm{ini}}\right)^{n} \nonumber \\
&+ \sum_{d\in \mathcal{S}_{\textrm{r}}} C^{\textrm{(r)}}(d,t) \left( \exp\left[ -\beta ( x - c_d )^2 \right] - \exp\left[ -\beta ( x_{\mathrm{ini}} - c_d )^2 \right] \right),
\end{align}
which means that it is enough to evaluate only one coefficient $C^{\textrm{(c)}}(n=0,T)$ to obtain $\varphi^{(m)}(x_\mathrm{ini},T)$. Since the Doi-Peliti method is abstract, there are several alternatives for its concrete representation. However, the above representation is natural in the sense that the target statistic is obtained with only one coefficient at the `origin' of the conventional basis.

To derive the time-evolution equation for the coefficients $\{C^{\textrm{(c)}}(n,t)\}$ and $\{C^{\textrm{(r)}}(d,t)\}$, it is necessary to specify the concrete form of the corresponding bra vectors; the inner products, such as in Eq.~\eqref{eq_inner_product_for_Z}, should be evaluated numerically. Here, we employ the following bra vectors:
\begin{align}
\langle n | = \int dx \, \delta(x-x_\mathrm{ini}) \left( \frac{d}{d(x-x_\mathrm{ini})} \right)^n (\cdot),
\label{eq_def_bra_vector_shift}
\end{align}
and 
\begin{align}
\langle \zeta_{0,d} | = \int dx \, \exp\left( - \beta ( x - c_d )^2 \right).
\end{align}
It is easy to confirm that the above bra vector in Eq.~\eqref{eq_def_bra_vector_shift} gives $\langle m | n \rangle = n! \delta_{m,n}$. As for $\langle n | \zeta_{0,d} \rangle$, $\langle \zeta_{0,d} | n \rangle$, and $\langle \zeta_{0,d'} | \zeta_{0,d} \rangle$, all the integrands are related to Gaussian distributions, and the numerical evaluation is not difficult; for example, it is possible to employ the numerical values obtained from the moment-generating function. Then, the matrices $Z$ and $L$ in Eq.~\eqref{eq_redundant_time_evolution_final} are numerically evaluated, and the time-evolution equation for the coefficients are derived.

For multivariate cases, we use
\begin{align}
\varphi^{(\bm{m})}(\bm{x},t)
= &\sum_{\bm{n} \in \mathcal{S}_{\textrm{c}}} C^{\textrm{(c)}}(\bm{n},t) \left(\bm{x} - \bm{x}_{\mathrm{ini}}\right)^{n} \nonumber \\
&+ \sum_{d\in \mathcal{S}_{\textrm{r}}}  C^{\textrm{(r)}}(d,t) \left( \exp\left[ -\beta \| \bm{x} - \bm{c}_d \|^2 \right] - \exp\left[ -\beta \| \bm{x}_{\mathrm{ini}} - \bm{c}_d \|^2 \right] \right).
\end{align}

As a demonstration, we consider the noisy version of the van der Pol equation \cite{VanderPol1926}:
\begin{align}
\begin{cases}
\mathrm{d}x_1 = x_2\mathrm{d}t + \nu_{11}\mathrm{d}W_1(t), \\
\mathrm{d}x_2 = (\epsilon x_2(1-x_1^2)-x_1)\mathrm{d}t + \nu_{22}\mathrm{d}W_2(t),
\end{cases}
\label{eq_noisy_van_der_pol}
\end{align}
where $\epsilon$ is a parameter, and $\nu_{11}$ and $\nu_{22}$ determine the noise amplitudes. This system is used as an example of nonlinear stochastic systems in some recent studies; see, for example, \cite{Crnjaric2020}. The corresponding time-evolution operator of the derived discrete state-process is
\begin{align}
\mathcal{L} = x_2\partial_{x_1}+ \epsilon x_2 \partial_{x_2} - \epsilon x_1^2 x_2 \partial_{x_2} - x_1 \partial_{x_2} 
+ \frac{1}{2}\nu_{11}^2\partial_{x_1}^2 + \frac{1}{2}\nu_{22}^2\partial_{x_2}^2,
\end{align}
and it is rewritten with the creation operator in Eq.~\eqref{eq_new_c} and the annihilation operator in Eq.~\eqref{eq_new_a} as follows:
\begin{align}
\mathcal{L} = & \left(a_2^\dagger a_1 + x_{2,\mathrm{ini}} a_1 \right)
+ \left( \epsilon a_2^\dagger a_2 + \epsilon x_{2,\mathrm{ini}} a_2\right) \nonumber \\
& + \left(- \epsilon a_1^\dagger a_1^\dagger a_2^\dagger a_2
- 2 \epsilon x_{1,\mathrm{ini}} a_1^\dagger a_2^\dagger a_2 - \epsilon x_{2,\mathrm{ini}} a_1^\dagger a_1^\dagger a_2 \right. \nonumber \\
&\quad \,\,\,\, \left.-2\epsilon x_{1,\mathrm{ini}} x_{2,\mathrm{ini}} a_1^\dagger a_2 - \epsilon x_{1,\mathrm{ini}}^2 a_2^\dagger - \epsilon x_{1,\mathrm{ini}}^2 x_{2,\mathrm{ini}} \right) \nonumber \\
&+ \left( a_1^\dagger a_2 - x_{1,\mathrm{ini}} a_2 \right) 
+ \frac{1}{2}\nu_{11}^2 a_1 a_1 + \frac{1}{2}\nu_{22}^2 a_2 a_2,
\end{align}
where $x_{1,\mathrm{ini}}$ and $x_{2,_\mathrm{ini}}$ are the elements of $\bm{x}_\mathrm{ini}$. Note that the creation and annihilation operators, $a_i^\dagger$ and $a_i$, act only on the $i$-th state. Next, we must specify the finite approximation of the basis, and the following two cases are considered:
\begin{itemize}
\item Conventional: 
$\mathcal{S}_\mathrm{c} = \{[0,0], [1,0], [0,1], [2,0], [1,1], [0,2], \cdots, [1,3], [0,4]\}, \mathcal{S}_\mathrm{r} = \emptyset$, i.e., the monomial basis up to $4$th order. This means that we employ the basis with $\{1, x_1, x_2, x_1^2, x_1 x_2, x_2^2, \cdots, x_1^1 x_2^3, x_2^4 \}$.
\item Redundant: $\mathcal{S}_\mathrm{c} = \{[0,0], [1,0], [0,1], [2,0], [1,1], [0,2]\}$ and $\mathcal{S}_\mathrm{r} = \{1,2,\dots, 9\}$. The total number of basis is the same as the above conventional one. The center vectors of the redundant basis, $\{\bm{c}_{d} | d = 1, \dots, 9\}$, are given in Appendix A.
\end{itemize}

\begin{figure}[tbp]
  \centering
  \includegraphics[width=9cm]{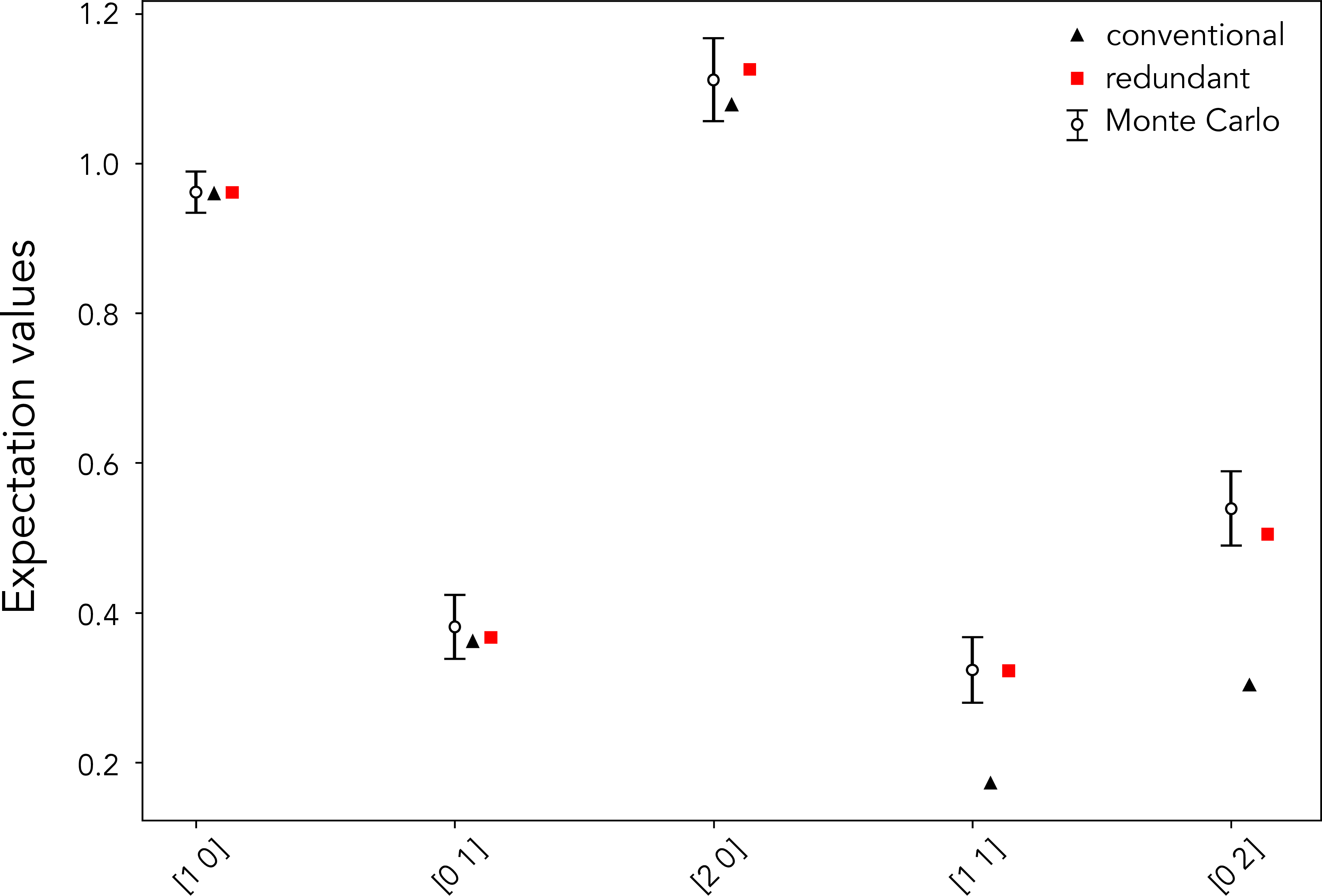}
  \caption{
Expectation values with the fixed initial condition $\bm{x}_\mathrm{ini}$. The horizontal axis represents the degree of the expectations; for example, $[1,1]$ implies the vertical axis value is $\mathbb{E}_T[(x_1 - x_{1, \mathrm{ini}})(x_2 - x_{2,\mathrm{ini}}) | \bm{x}(t=0) = \bm{x}_\mathrm{ini}]$. The black triangles and the red squares correspond to the statistics calculated from the conventional and redundant basis, respectively. The filled circles depict the values obtained from the Monte Carlo calculations. The error bars represent the standard deviations.
}
  \label{fig:experiment_result}
\end{figure}

The numerical details are denoted in Appendix A. Figure~\ref{fig:experiment_result} shows the results of the numerical experiments. As written above, the initial condition of the derived discrete-state process corresponds to the statistic to be evaluated in the original stochastic differential equation. Figure~\ref{fig:experiment_result} depicts the expectations up to the second order. The horizontal axis represents the degree of the expectations; for example, $[1,1]$ implies the vertical axis value is $\mathbb{E}_T[(x_1 - x_{1, \mathrm{ini}})(x_2 - x_{2,\mathrm{ini}}) | \bm{x}(t=0) = \bm{x}_\mathrm{ini}]$. The triangles and squares in Fig.~\ref{fig:experiment_result} correspond to the conditional expectation values obtained from the conventional and redundant bases, respectively. The circles represent the results from the Monte Carlo simulations; the error bars show the standard deviations. We find that $\mathbb{E}_T[(x_1 - x_{1,\mathrm{ini}})(x_2 - x_{2,\mathrm{ini}})| \bm{x}(t=0) = \bm{x}_\mathrm{ini}]$ and $\mathbb{E}_T[(x_2 - x_{2,\mathrm{ini}})^2| \bm{x}(t=0) = \bm{x}_\mathrm{ini}]$ obtained from the conventional basis are far from the Monte Carlo values. By contrast, all the evaluated values from the redundant basis are close to the Monte Carlo results. This numerical demonstration shows that a derived discrete-state process captures the essence of the original stochastic differential equation adequately compared with the conventional basis.

\section{Conclusions}

In the present work, we proposed the use of the redundant basis for the Doi-Peliti method and showed an application. The results showed that the redundant basis yields a complicated discrete-state process with long-range hopping. This feature differs from a discrete-state process derived from the conventional one. Instead of the redundancy and the complexity, the derived discrete-state process captured the essential part of the original stochastic differential equation; the derived discrete-state process yields accurate estimations for the statistics compared with the conventional discrete-state process with the same number of states. To derive the discrete-state process from the redundant basis, we used the interpretation of the Doi-Peliti method by means of the generating function approach. In addition, the inverse matrix built by the inner product of the basis is necessary for the derivation.

There are some remaining comments.

As commented on Sect.~\ref{sec_review_dual}, there are several interpretations of the Doi-Peliti method. In the current work, we only use this feature for the shift of origin in the function expansion. As discussed in \cite{Ohkubo2021}, the shift is crucial to employ an algorithm based on combinatorics. The aim of the present paper is to show the following fact: The usage of redundant basis yields discrete-state processes with long-range hopping, which are different from previous ones. Since we employed only the limited interpretation, studies combined with the various interpretation remain as future works. In addition, the interpretations are deeply related to the quantum probability theory or algebraic probability theory \cite{Hora_book}. The mathematical discussions for the interpretations are hopeful for the future.

As for the redundant basis function, we employed the Gaussian functions because of the computational reason; it is easy to deal with the integrals for the Gaussian functions. Of course, other types of redundant basis functions are also possible in principle. Computational aspects are beyond the scope of the present paper, and we hope that the current work stimulates researchers in computational physics. There may be more appropriate redundant basis functions.

We demonstrated that the redundant basis functions yield better numerical estimates than the conventional ones. Estimations for higher-order moments are also possible; we have checked estimations up to the 4th order moments. It is hopeful to discuss the reason in the future. There is a conjecture; the discrete-state process derived from the redundant basis functions could use the state space efficiently because of the long-range hopping as depicted in Fig.~\ref{fig_discrete_process_from_redundant_basis}. However, the discrete-state process is a little complicated; it is difficult to obtain physical insights. It will be crucial to seek suitable systems and redundant basis functions to investigate the physical meanings.

It will be also hopeful to investigate what kinds of basis functions are crucial for the approximation. The use of higher-order monomials may provide hints on the choice of redundant basis functions since we would obtain information on sufficient nonlinearity for approximation. In preliminary numerical experiments, we confirmed that monomials up to the 5th order yield better estimates for the numerical experiments in Sect.~\ref{subsec_application}. We also performed preliminary numerical experiments for different systems, including the Ornstein-Uhlenbeck process and the noisy Duffing equation without external forces. However, when we deal with stochastic differential equations with high dimensions, the number of monomial basis functions increases rapidly. Recently, several works employed tensor networks or tensor trains to investigate chemical master equations (for example, see \cite{Gelss_thesis, Ion2021, Nicholson2023}). The tensor networks will be beneficial to seek the essential redundant basis functions in the future.

The current work succeeded in greatly expanding the range of derived discrete-state processes. We hope the derivation techniques here will help future research on stochastic processes.

\backmatter


\bmhead{Acknowledgments}

This work was supported by JST FOREST Program (Grant Number JPMJFR216K, Japan). 



\begin{appendices}

\section{Numerical details}

For the noisy van der Pol equation, we here use the parameters $\epsilon = 1.0$ and $\nu_{11} = \nu_{22} = 0.5$. The initial state is $\bm{x}_\mathrm{ini} = (-0.1, 0.9)^\mathrm{T}$, and the statistics at $T=0.8$ are evaluated.

In the Monte Carlo simulations, we take the averages of $200$ samples to evaluate the mean values of the statistics. We repeat this procedure $200$ times to obtain the standard deviation of the means. The time-discretization with $\Delta t = 10^{-3}$ is used in the Euler-Maruyama approximation \cite{Kloeden1992}. 

For the conventional basis, the number of basis (states) is $15$, as denoted in Sect.~\ref{sec_application}. We must specify the center positions $\{\bm{c}_d | d = 1, \dots, 9\}$ for the redundant basis $| \zeta_{0,d}\rangle$. First, we solve the van der Pol equation without noise, i.e., $\nu_{11} = \nu_{22} = 0$. Second, we use equally-spaced $25\times 25 = 625$ grid points in the range $[-2,2]$ on each dimension. Then, we tentatively set the center position $\bm{c}_d$ on each grid points, and the values of $\exp(-\beta \|\bm{x}(T)-{c}_d\|^2)$ are numerically evaluated. Third, we select $9$ grid points with the largest values; the selected points consist of the set $\mathcal{S}_\mathrm{r}$. Table~\ref{table_centers} shows the selected center points. In the time-evolution of the derived discrete-state process, we use $\beta = 4$ in $| \zeta_{0,d}\rangle$.

\begin{table}[tbp] 
\caption{The center points used as $\bm{c}_d$ in $| \zeta_{0,d}\rangle$.}
\label{table_centers}
\begin{tabular}{cl}
\toprule
$d$ & $\bm{c}_d = (x_1, x_2)$ \\
\midrule
1 & $(2/3, 4/3)$\\
2 & $(2/3, 3/2)$\\
3 & $(5/6, 7/6)$\\
4 & $(5/6, 4/3)$\\
5 & $(5/6, 3/2)$\\
6 & $(1, 7/6)$\\
7 & $(1, 4/3)$\\
8 & $(1, 3/2)$\\
9 & $(7/6, 4/3)$ \\ 
\botrule
\end{tabular}
\end{table}

\end{appendices}

\end{document}